# Quantitative Lattice Simulations of the Dense Amorphous Phase in Semicrystalline Polymers : Size and Energy Effects in 2-D Lattices

Joydeep Mukherjee[1]
Antony N. Beris[2, *]
[1] The Dow Chemical Company, Freeport, TX -77541
[2]Department of Chemical Engineering, University of Delaware, Newark, DE-19716;
Email: beris@udel.edu, Phone: +1(302) 831 8018, Fax: +1(302) 831 1048
[*]Author to whom correspondence should be addressed

## Summary

In this work we illustrate our novel quantitative simulation approach for dense amorphous polymer systems, as discussed in our previous work[Kulkarni et al., A Novel Approach for Lattice Simulations of Polymer Chains in Dense Amorphous Polymer Systems: Method Development and Validation with 2-D Lattices, arXiV, 2008] in applications involving large lattice sizes and high energetic bias. We first demonstrate how the topology of the microstate ensemble in 2-D lattices presents a serious challenge for the collection of accurate and reliable quantitative results (i.e., with simultaneous determination of error bars) for large lattices. This necessitates a further enhancement of our Monte Carlo simulation scheme to sample effectively a meaningful 2-D lattice configurational subspace. Two techniques were investigated: simulated annealing and parallel tempering, to avoid trapping near a local free energy minimum in simulations at high energetic bias. Extensive results of the prediction of various chain conformation statistics and thermodynamic quantities, in the thermodynamic limit (i.e. infinite lateral sizes) are presented.

## 1 Introduction

In our previous paper[1], we proposed and validated a new quantitative and robust lattice-based methodology for investigating dense polymer chains. The discussion there was limited to small size 2-D lattices (up to 8x8 nodes) for the purposes of validation of the stochastic scheme against exact results. However, in order to get quantitative estimates of conformational statistics and thermodynamic variables in the thermodynamic limit (i.e. in the limit of infinite lateral extent of the lattice), larger lateral dimension lattices need to be used. These larger sizes are necessary so that an extrapolation to the limit of infinite lateral dimension can be performed and thus obtain mesh-independent answers.

Furthermore, investigations at high energetic bias (in the present case arising from energetic penalties on intra-chain tight folds) are also of critical importance. This poses additional computational challenges because at high energetic biases, trapping into local minima in the potential energy surface is possible[2-5].

Traditional Monte Carlo methods typically suffer in sampling efficiently at high energetic bias leading to results that are dependent on the initial conditions with systematic errors and unreliable error estimates[4-7]. To alleviate this problem we have implemented two different enhancements of our Monte Carlo scheme in addition to the reweighting Monte Carlo methods discussed in our previous work which was further modified in order to apply it to large lattice sizes, as discussed in section 2. These are simulated annealing, discussed in section 3 and parallel tempering, discussed in section 4. In section 5, predictions obtained from these schemes on some important conformational statistics and thermodynamic quantities are presented in the thermodynamic limit. Finally, our conclusions follow in section 6.

## 2 Monte Carlo Sampling at Large Sizes and High Energetics: The Dilemma with Reweighed Averaging

As mentioned before, an inherent problem encountered in all traditional Monte Carlo implementations is that the scheme tends to get trapped into a local minimum at high energetics (low temperatures). This leads to systematic errors due to misrepresentation of the contribution from the states corresponding to higher internal energy. We have seen this with our Monte Carlo scheme even in small 8x8 node lattice simulations at high energetic bias as presented in our previous work[1]. Under those conditions the normal Monte Carlo scheme fails to provide consistent averages and meaningful error estimates for various chain statistics. The solution proposed there, and successfully implemented in 8x8 node lattice was reweighting[2], i.e. calculating the various statistics corresponding to higher energetics by using the Monte Carlo sampling at a lower energy and reweighting the statistics in a post-processing step. However the success of reweighting scheme relies heavily on the capability to sample a substantial portion of the total number of microstates at lower energies so as to make sure that the most energetic contributions have been sampled in a fashion representative of their total population. Whereas this is feasible for a relatively small population size in case of an 8x8 node lattice (having about 12.8x10$^9$ microstates, with one perfectly crystalline state corresponding to zero tight folds and 912 microstates corresponding to two tight folds.), this already becomes infeasible with an 8x16 size lattice.

Here we describe in detail our experience with an 8x16 node lattice. The stochastic enumeration scheme[1] was first used to estimate the total number of microstates for an 8x16 node lattices. The value predicted is approximately 6.185x10$^{19}$! As such, even a relatively large sample size of 32 billion which is close to the limit that can be generated within a feasible amount of time (about a week) on an Athlon MP 2000+ dual processor system, represents at best only a small portion of the total population. Twenty such calculations were carried out on a 50-processor Beowulf cluster using different seeds to random number generator. Figure 1 shows a comparison of the results of one key conformation statistics, the fraction of chains forming tight folds in the first layer, as a function of the energetic penalties on tight folds. The predictions at high energetic bias ($E_\eta > 1$) were obtained by reweighting the distribution obtained at a base energy of 1. First focus on the raw (uncorrected) results represented by the two runs with seeds to random number generator 2 and 8 (represented by cross and filled triangle in Figure 1). It is obvious from these results that even the reweighting scheme fails to provide quantitative estimates in this case, as there is a wide variation in the predictions with large systematic errors and meaningless a priori error estimates.

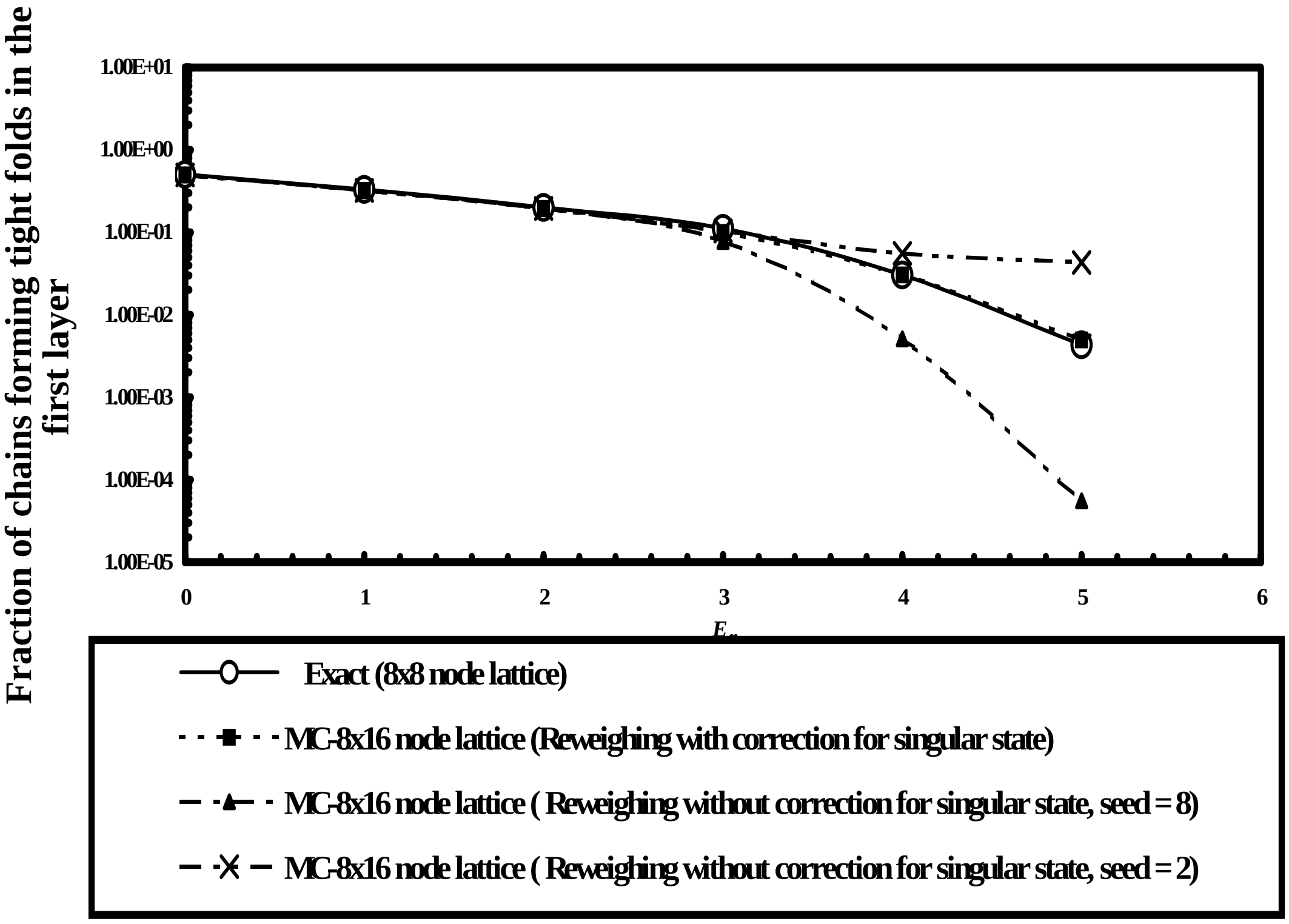

**Figure 1.** Prediction of fraction of chains forming tight folds in the first layer of an 8x16 node lattice using various methods and its comparison with the exact values for 8x8 node lattice.

To reason out this discrepancy, the frequency distribution of the samples generated in these cases was analyzed with respect to total number of tight folds. As shown in Figure 2, there is a huge difference in the frequency distribution especially with respect to structures with low number of tight folds. However, given the presence of reweighting, the structures with less number of tight folds (corresponding to lesser energy) have much higher weights at high energetic penalties on tight folds and correspondingly they influence the final average to a great extent. Any misrepresentation of their frequency (due to Monte Carlo sampling) might result in the calculation of a biased average shifted towards the statistics corresponding to structures with lesser or higher number of tight folds depending on the fluctuation. Hence, when using a seed value of 8 where the structures with zero tight folds are represented with a high frequency, we get as a result, a low estimate of the fraction of chains forming tight folds in the first layer as seen in Figure 1. The opposite is true in the case of seed value of 2.

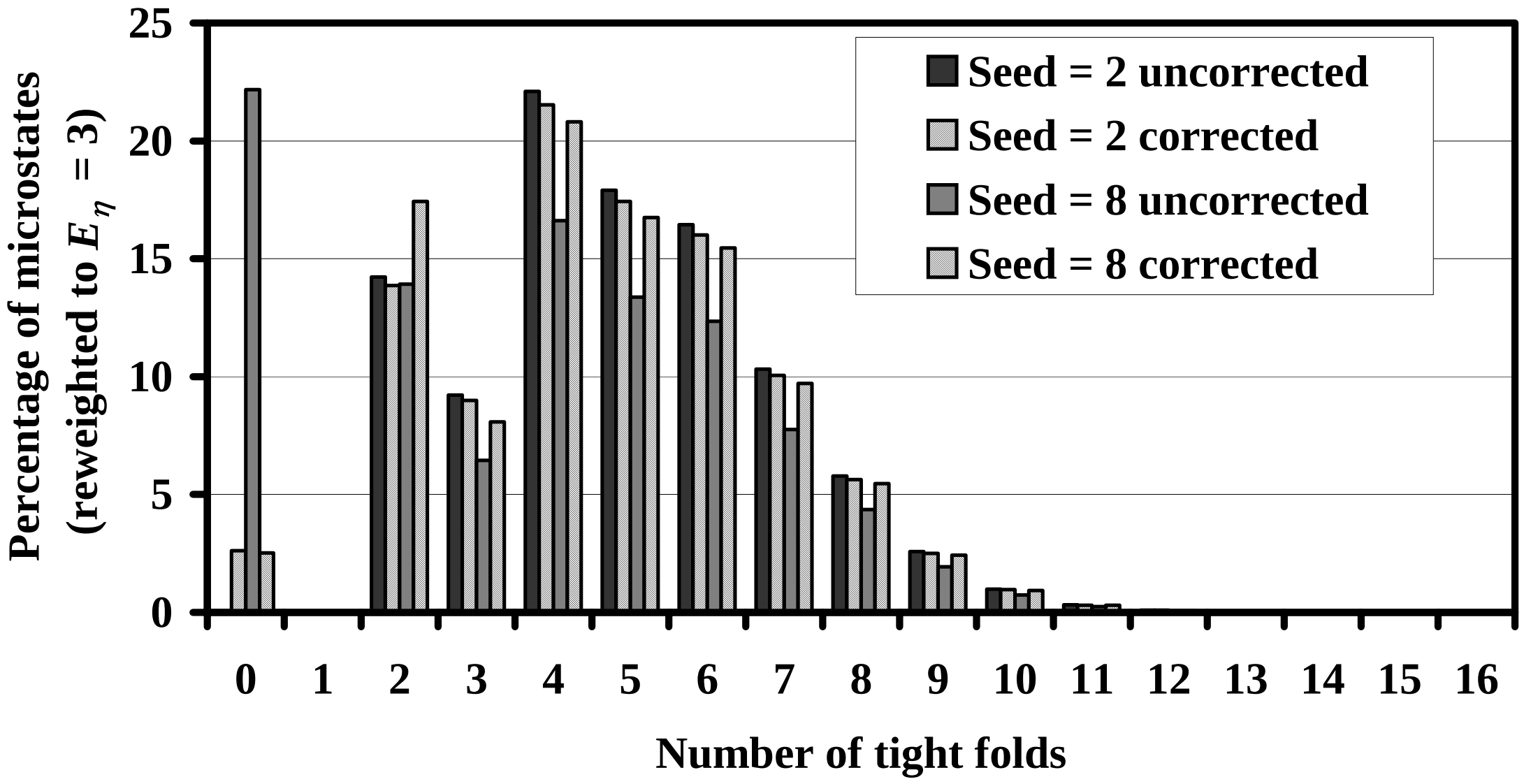

**Figure 2.** Comparison of frequency distribution of visiting a microstate with respect to the total number of tight folds for two different seeds to the random number generator.

In general, such a discrepancy is an inherent feature of reweighting scheme and cannot be corrected other than by unreasonably increasing the sample size. However for this particular case the state that is responsible for causing this problem is primarily only one and known a priori- the zero fold perfectly crystalline state. Since (through the use of stochastic enumeration), we know its relative contribution for zero energetics (1 out of $6.185\text{x}10^{19}$), it is feasible to calculate its anticipated frequency in the 32 billion sample at energetics of $E_\eta$ =1 and therefore correct the frequencies of this and all other states (based on normalization condition). When this is done the skewed frequencies distribution of states is corrected as shown in Figure 1. When this corrected distribution was used to calculate the new averages those come fairly close with the exact predictions for 8x8 node lattice as is shown in Figure 1. The error bars shown in Figure 1 represent two standard deviations about the averages and in some cases cannot be seen at the scale of the figure.

The technique outlined here could be implemented in this case since we had a good estimate of the population size for the 8x16 node lattice and only one known state was primarily responsible in altering the frequency count. Even so, and for this lattice size (8x16), the computational requirements in order to obtain quantitative estimates have been substantial (8 days with 20 CPUs running simultaneously, in other words 160 CPU days!). The situation becomes even worse for larger lattices sizes as the population size increases exponentially and a larger sample size than 32 billion would have been required that poses prohibitively high computational time requirements. So alternative tools to investigate larger lattice sizes at high energetic bias need to be investigated. In the next sections we present two such special techniques, simulated annealing and parallel tempering.

## 3 Simulated Annealing

### 3.1 Description of method and implementation

The simulated annealing scheme attempts to avoid local minimum trapping by obtaining an average over carefully tailored Monte Carlo sequences with initial guesses generated systematically from a sequence of Monte Carlo runs at different energetics. The annealing procedure (first increasing and then decreasing the temperature in a gradual fashion) followed here is the one that we found most efficient and is described briefly:

1. Starting from the current guess (crystalline state initially), increase the temperature of simulation to a desired level (corresponding to $E_\eta = E_0$) and allow the Monte Carlo process to equilibrate for some predetermined number of steps.
2. Randomly decide to increase or decrease the energy parameter $E_\eta$ by a pre-specified amount $\Delta\varepsilon$ (typically 0.5) with a certain probability such that if there had been a similar change in the pervious step, there is a bias in favor of a change in the opposite direction (say 2/3 versus 1/3 probabilities) provided $E_\eta$ is not smaller than the minimum $E_\eta$ (equal to zero here) and vice versa. Then run the Monte Carlo steps at the new energetics allowing the results to equilibrate for some predetermined Monte Carlo steps.
3. Continue the annealing process until the desired energetics is reached. Equilibrate and collect data for statistics for a number of Monte Carlo steps. Repeat the annealing process starting with step 1.

The above procedure allows for the annealing to be gradual and relaxed.

### 3.2 Initial results: Local minima trapping

The typical parameters used in simulated annealing were: $E_0 = 1.0$, number of equilibration steps = maximum (1000, $E_\eta$ x 1000). The number of equilibration steps was decided based on the correlation length of sampling at each $E_\eta$, allowing enough steps for the Monte Carlo sampling to remove any memory of the variation in temperature. Figure 3 shows a simulated annealing prediction for the fraction of chains forming tight folds in first layer for an 8x16 node lattice. To allow an interpretation, the results are presented separately for each Monte Carlo sampling following the simulated annealing procedure outlined in the previous section. The results clearly indicate that there are two separate populations that are fairly well separated by an energy barrier that is difficult to cross. The bottom one represents the global minimum and is more probable than the top one which is a secondary minimum. This figure depicts the local trapping of the scheme at such minima very clearly. Note that due to the random selection based on the annealing process the overall average considering both the global and secondary minima does not yield the thermodynamic average but rather some intermediate average, which varies widely with the number of jumps, made from one population to the other.

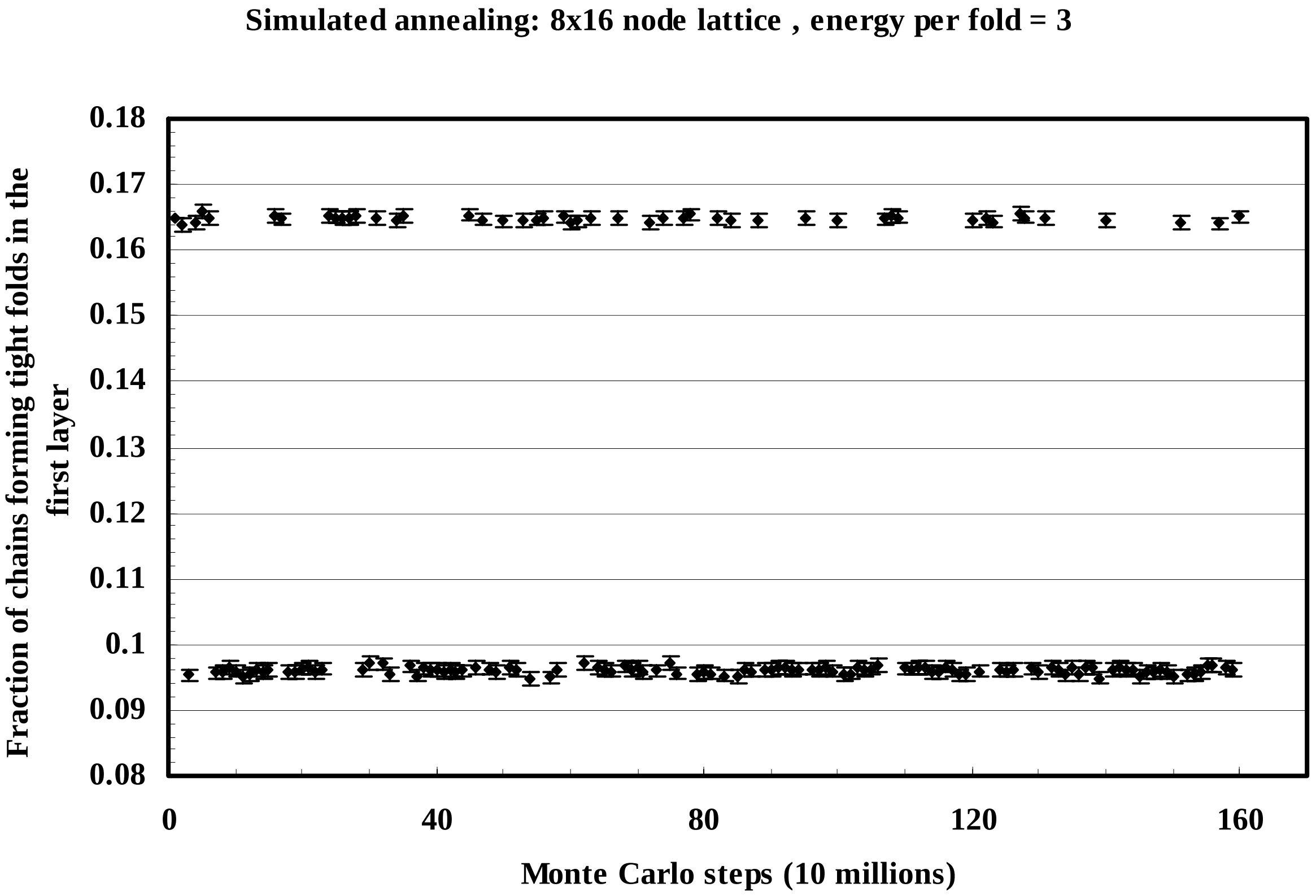

**Figure 3.** Simulated annealing predictions for fraction of chains forming tight folds in first layer for an 8x16 node lattice.

Further insight can be provided by a comparison of the frequency distribution of the sampled microstates with respect to total number of tight folds corresponding to either one of the two populations sampled, as shown in Figure 4. We also show the “correct” distribution based on the corrected reweighting process discussed in section 2. The figure clearly indicates that while population 1 (corresponding to the region near the global minimum) has states with less number of tight folds “over sampled” compared to the correct distribution, the opposite is observed with population 2 (corresponding to the secondary minimum). If however the distribution of population is assigned a weight of 78% and the population 2 is assigned a weight of 22% and a weighted average of the two populations is calculated, the weighted distribution is fairly close to the correct distribution corresponding to the thermodynamic population. This is encouraging as it indicates that the scheme does capture eventually all the population but in two separate ensembles that unfortunately need to be corrected a posteriori. Efforts were then made to identify the key features that separate the two populations and these are discussed next.

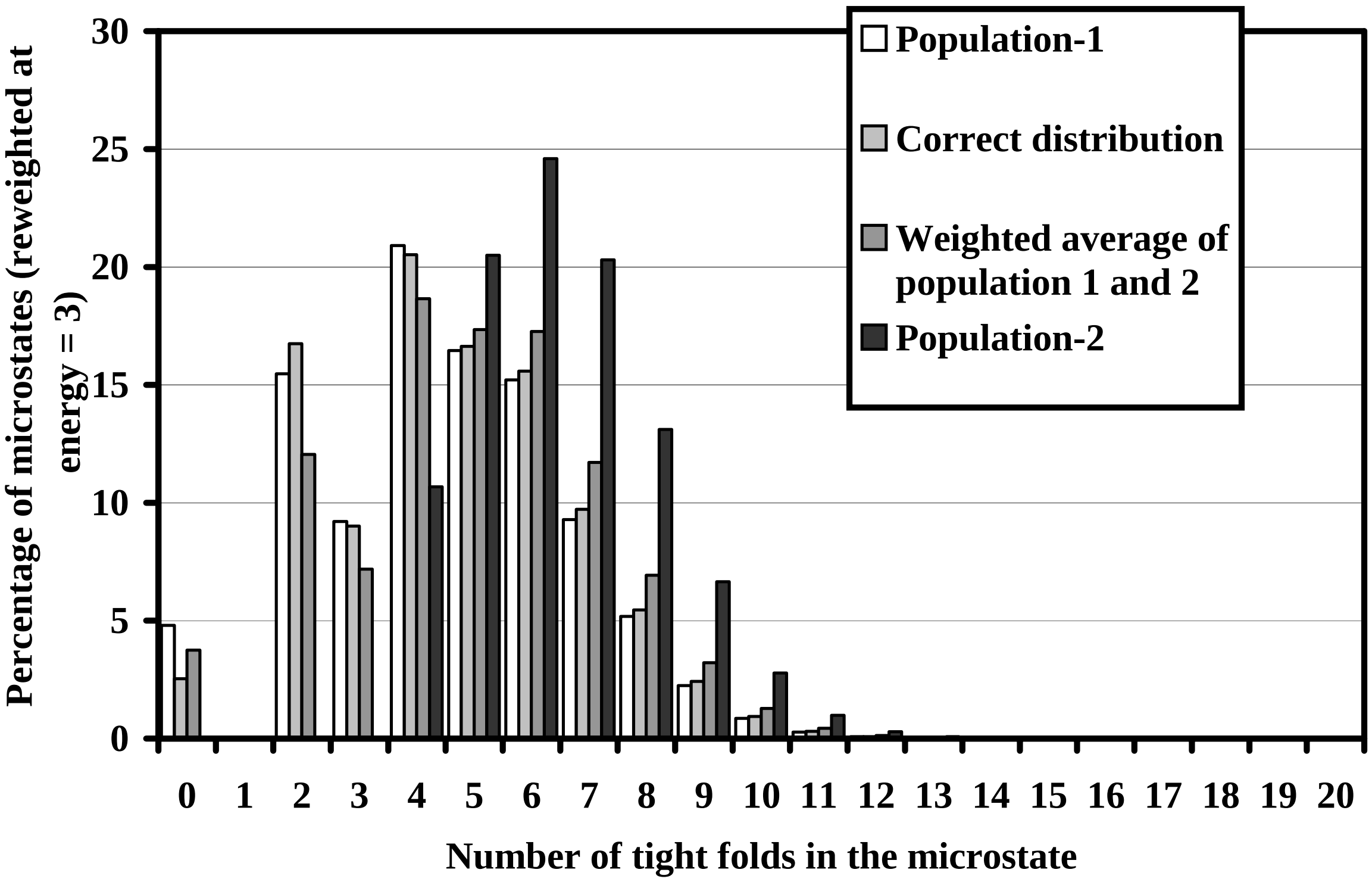

**Figure 4.** Comparison of frequency distribution of visiting a microstate with respect to the total number of tight folds for the two minima (population 1: global minimum, population 2: secondary minimum), the correct distribution corresponding to the thermodynamic population and weighted distribution of population 1 and 2.

### 3.2 Interpretation of the results based on investigations of the topology of the 2-D lattice structures

The observations in the previous section necessitated further investigations into the key features that separate the two populations. Sample 8x16 node lattice structures corresponding to each of the two populations were plotted. Typical sample results belonging to either one of the two populations are shown in Figure 5. As it is seen there, the key difference is that population 1 had bridges (chains connecting the two crystalline boundaries) that have exclusively an even number of horizontal segments while the structures in population 2 had an odd number.

Indeed it is easy to show that topological constraints of 2-D lattice representation result, for an even number of lattice nodes (16 here) along the horizontal direction and the imposition of periodic boundary conditions, in the impossibility of having microstates with both kinds of bridges present. We either have microstates with no or all bridges with even number of horizontal segments (those seen in population 1) or all bridges having an odd number of horizontal segments, those seen in population 2. So these two populations are structurally separate and it is remarkable that the existing algorithm allows sampling both of them at low energetics (this is attributed to the possibility of effecting global changes by changing the submicrostates of all sublattices simultaneously). However such a transition becomes unlikely at high energetics leading to trapping in local energy minima. This analysis also explains the reason behind the fact that the population 1 has the lowest energy. Since population 1 has structures with bridges having even number of horizontal segments, it is the population that contains the perfect crystalline state (a structure that has all bridges with no horizontal segments).

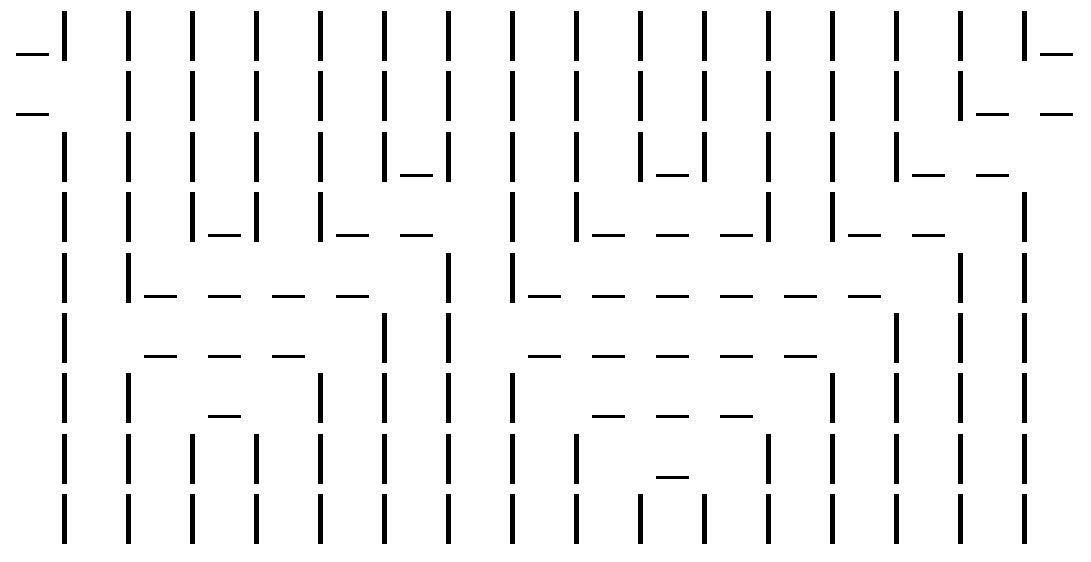

Population 1: Bridges have even horizontal segments

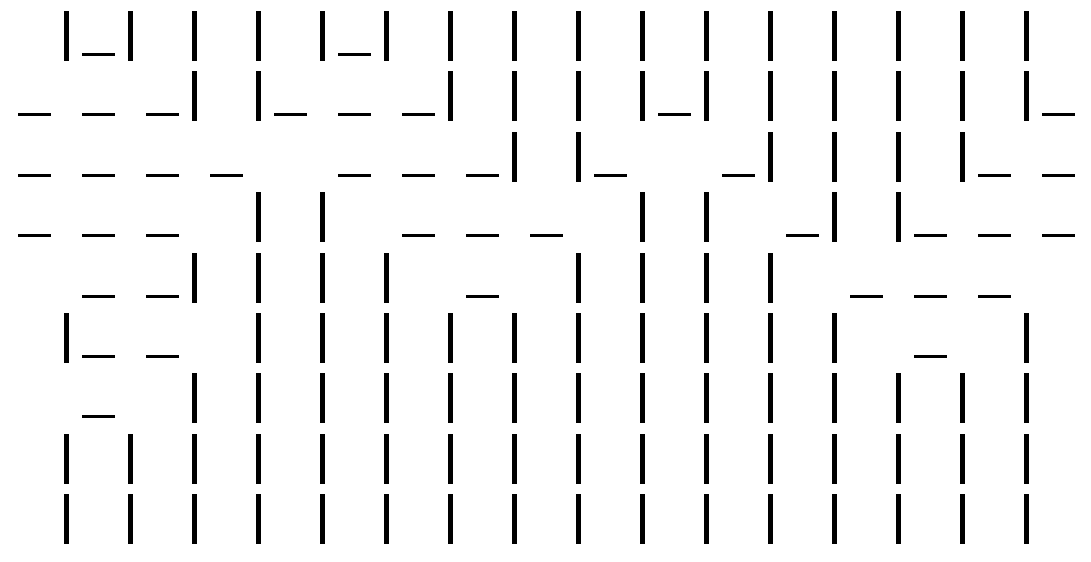

Population 2 : Bridges have odd horizontal

**Figure 5**. 8x16 node lattice structures corresponding to populations 1 and 2.

As the energetic penalty on tight folds is increased the sampling is more biased towards structures containing less and less number of tight folds and the perfect crystalline state is the most favored structure. However, as a result of the annealing process, occasionally the sampling process ends and gets trapped into the population 2 (which in an average sense corresponds to a relatively higher energy) and remains there until another annealing step is performed. Thus the observation of two separate populations at high energetic penalties is more a topological feature characteristic of 2-D lattices resulting from the lattice filling restrictions. The fact is that the two populations are topologically distinct, not allowing (in the thermodynamic limit) a transition from one to the other based solely on local changes. In the limit of infinite lateral size, an infinite number of changes are necessary. Thus it makes more sense to study only one of the two populations. Since the one with the even number of horizontal segments in the bridges is the one containing both interesting limits (with no or all bridges) and that with the smaller energy, it makes sense to use population 1 for our studies. This is an additional constraint that we implemented consistently in our scheme sampling structures belonging to exclusively to population 1. This ended up producing always reliable and consistent results with consistent error estimates and provided a reliable estimate of the quantities under the abovementioned restriction, as shown in Figure 7 below, in which the error bars (which show 2 standard deviations about the average) are seen to consistently decrease in size with CPU time. In that figure we show a comparison against the results obtained with the parallel tempering procedure discussed next.

## 4 Parallel Tempering

### 4.1 Method description

The basic idea behind parallel tempering[3] is to perform several different simulations simultaneously at different temperatures (i.e. different energetic penalties on tight folds) spanning a range of values. Every so often one swaps the states of the system between two of the simulations running at two adjacent values of energetic parameters (temperatures) with a certain probability so that the Boltzmann distribution at each simulation still holds. Such swapping of states allows one to get over energetic barriers at each simulation while sampling the overall population with the correct probabilities. The parallel tempering scheme algorithm is described briefly below[3]:

1. Start N Monte Carlo simulations at different energy per fold penalties $E_1 < E_2 < E_3 .... E_N$, (corresponding to different temperatures $T_1 > T_2 > T_3 .... > T_N$). This can be done most efficiently in parallel, running on N processors: $P_1, P_2 .... P_N$.
2. Each processor $P_i$ (i < N) initiates a request (in an asynchronous fashion) to swap states with the simulation running at $P_{i+1}$.
3. Every so often (with a predetermined frequency depending on the precalculated correlation length of the Monte Carlo steps), the simulation running on processor $P_i$ decides to swap its state with the state of simulation running on processor, $P_{i-1}$, (i.e. one running at immediately higher temperature or lower $E_\eta$). This request is accepted by processor $P_{i-1}$ with an acceptance probability, $A$:

$$A = \begin{cases} e^{(E_i - E_{i-1}) \times \Delta N} & \text{if } \Delta N < 0 \\ 1 & \text{if } \Delta N > 0 \end{cases}$$

where

$$\Delta N = N_{tightfolds_i} - N_{tightfolds_{i-1}}$$

represents the difference in the number of tight folds in the states to be exchanged between the two simulations.

4. $P_{i-1}$ initiates a new request (in an asynchronous fashion) to swap states with simulation running at $P_i$ and the simulation continues until a predetermined number of samples are generated.

According to discussion in the previous section, we implemented the parallel tempering scheme with the constraint of sampling population 1 states only.

### 4.2 Discussion of performance

A key factor in the fine-tuning of the performance of the annealing scheme is the frequency with which a swapping of states is proposed. This frequency scales with the correlation length of the Monte Carlo sampling scheme and is decided heuristically. Here we chose to use a swapping move proposal frequency of 4 times the correlation length as it yielded a fast decrease in error with CPU time. Another key factor in the performance is deciding the different energetic penalty per folds (or temperatures) at which the simulations must be carried out so as to allow a reasonable swapping of states between simulations. This is very necessary for the efficiency of the scheme. This is done heuristically too in such a way so that there is considerable overlap in the energies sampled. Here, for simplicity, we chose to run the parallel tempering scheme with energetic penalties per fold differing by 0.5 between adjacent simulations. In Figure 6, a histogram of the frequencies with which various microstates are sampled with respect to the number of tight folds in them is shown for three adjacent energetic parameter value, which is indicative of the tight fold energy surface spanned for an 8 x16 node lattice. As Figure 6 clearly shows there is a sufficient overlap between the energy surfaces of different simulations (shown here are simulations run at energetic penalties of 3.0, 3.5 and 4.0). This ensures a favorable probability of success for a swapping move.

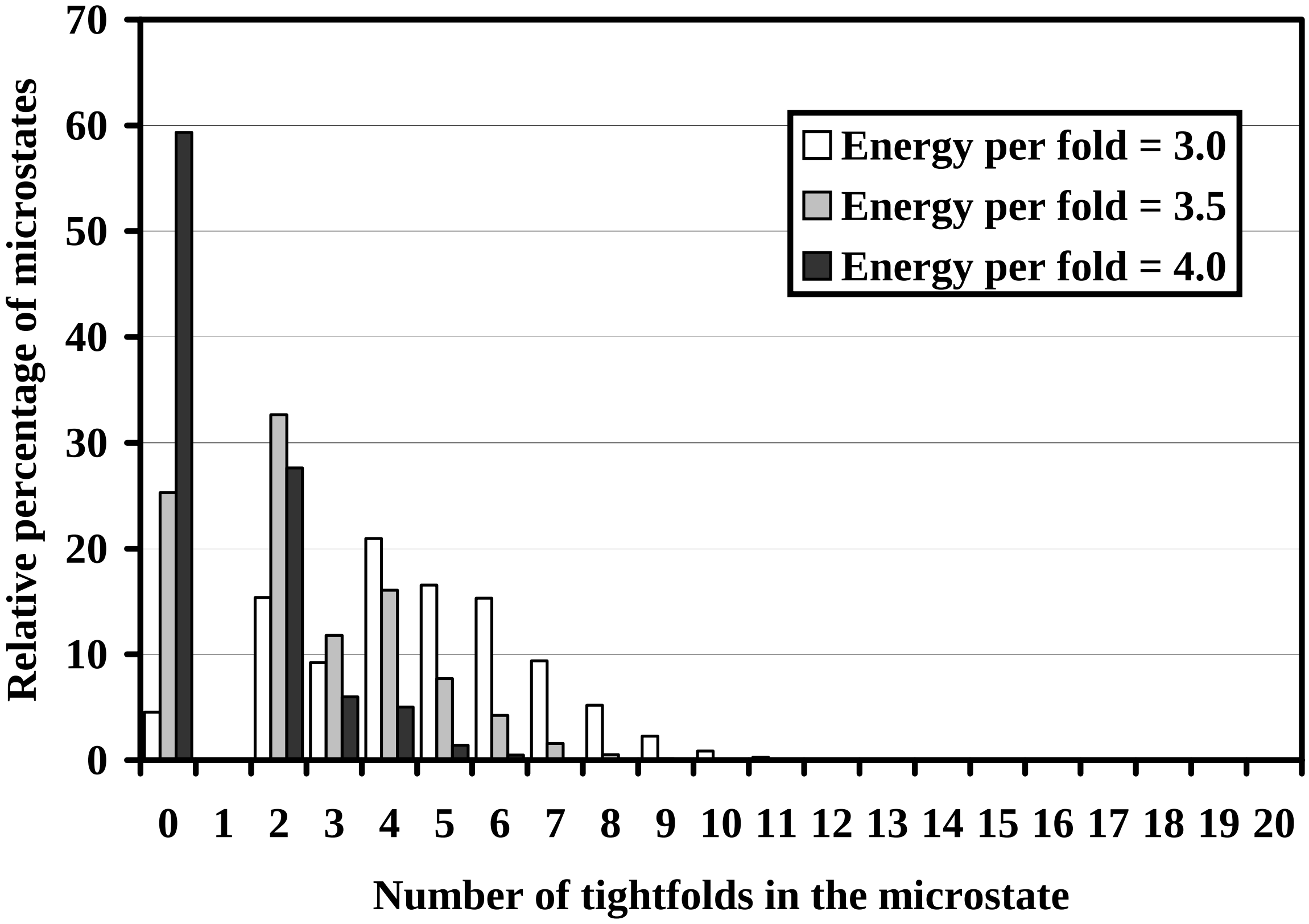

**Figure 6.** Energy spectrum of different simulations.

After fine-tuning the frequency and energetic parameter values, the performance of the scheme was compared with the simulated annealing scheme. A comparison against respect to prediction of the fraction of chains forming tight folds is shown in Figure 7. The running average and standard deviation is plotted as a function of the CPU time requirements. It is seen that the predictions of both the schemes are self and mutually consistent (i.e. resulting in overlapping confidence intervals).

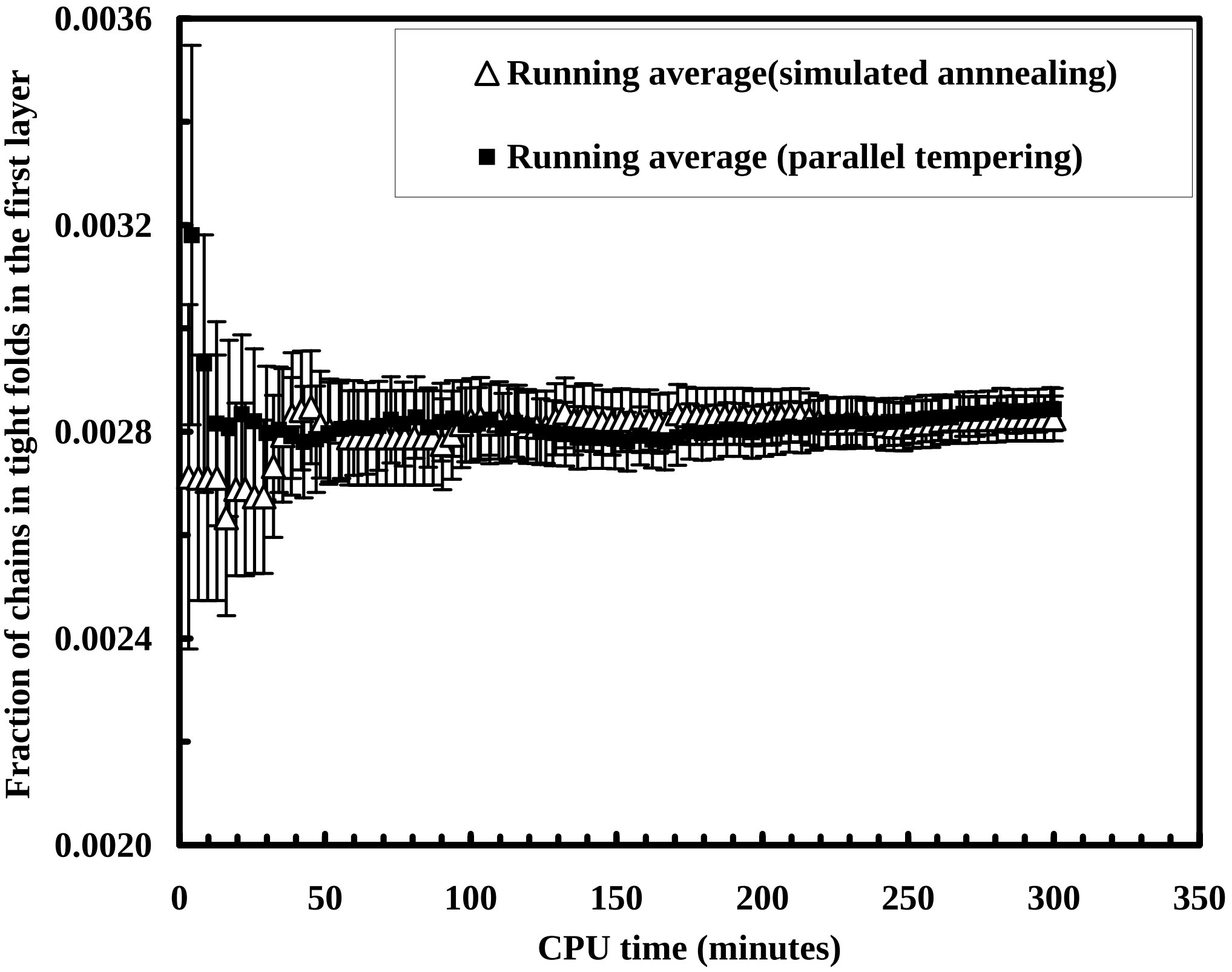

**Figure 7.** Comparison of results pertaining to fraction of folds forming tight folds in the first layer for 8x16 node lattice from two approaches: simulated annealing and parallel tempering.

In particular, we see from Figure 7 that the decrease in error with time is almost the same for both schemes. Although there performance of parallel tempering scheme seems at best almost similar to the simulated annealing scheme and one doesn't gain much as far as the efficiency of sampling is concerned, it is still a very useful scheme as the parallel nature allows several simulations to be run simultaneously yielding results at different energies simultaneously which is more tedious to achieve in a serial computing structure.

## 5 Results and Discussion

In this section we discuss the capability of our Monte Carlo scheme to achieve quantitatively robust predictions of various important conformational and thermodynamic quantities in the thermodynamic limit.

### 5.1 Estimation of amorphous chain conformational statistics

The most pertinent conformational quantities of interest for an amorphous dense assembly of chains in a lattice are: a) the order parameter, b) the length of loops, c) the length of bridges, d) the fraction of chains forming loops, e) the fraction of chains forming bridges. The order parameter is of critical importance and can be considered as a measure of the degree of orientation in the sandwiched amorphous region. It is defined as the fraction of chain segments in a row that are oriented along the crystal axis. Therefore, in our lattice where the vertical direction is the crystal axis, the order parameter for any row "*i*", is defined as $S_i = 2\langle x_i \rangle - 1$, where $x_i$ is the fraction of segments exiting row *i* and oriented in the vertical direction. So it takes a value of 1 in the perfect crystal and 0 in the completely amorphous phase.

As mentioned before, it is necessary to estimate these conformational statistics in the thermodynamic limit ($N_0 \rightarrow \infty$) so as to relax the periodicity boundary conditions along the lateral dimension. This is done here by evaluating conformational statistics of lattices with varying number of lattice nodes in the lateral direction, $N_0$, (8 nodes, 16 nodes, 24 nodes and 32 nodes). Then, plotting the quantities of interest with respect to $1/N_0$ and extrapolating to zero yields the desired estimate. The procedure is illustrated here in Figure 8. Figure 8 shows the estimation of the order parameter for different rows for the case of $E_\eta = 3$ in the thermodynamic limit. The variation in the order parameter is seen to be almost insignificant after a lattice lateral length of 24 or 32 unit segments. So we can conclude that for a lattice with interlamellar spacing of 8 unit segments, the lateral dimension of 32 units yields the estimates of quantities in the thermodynamic limit. It is important to note that the error bars (which represent twice the standard deviation) in these figures are too small to be noticed. This is extremely important if any meaningful extrapolation of the data is to be performed.

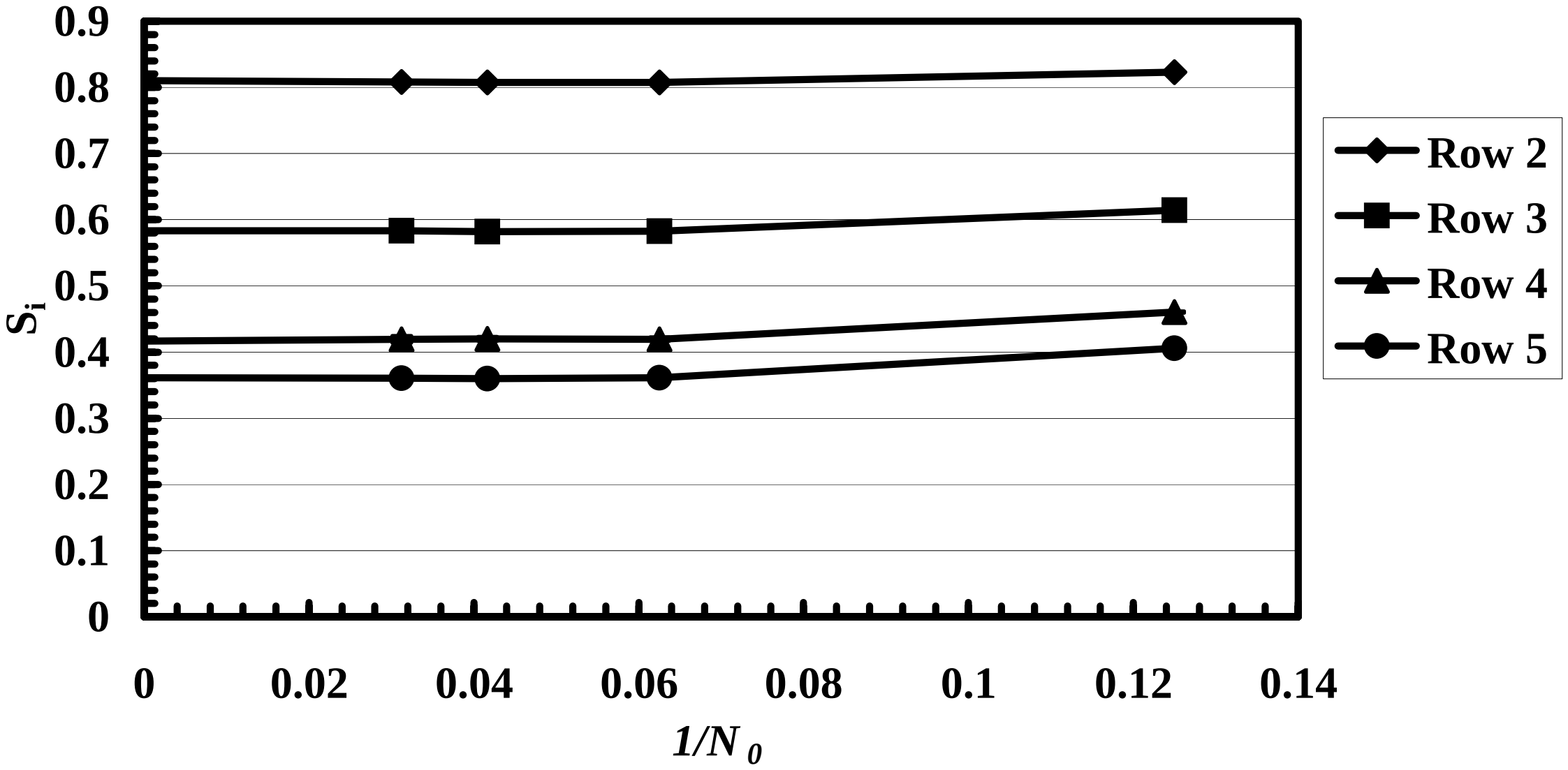

**Figure 8.** Estimation the order parameter for different rows in the thermodynamic limit for the case of $E_\eta = 3$.

The order parameter (estimated at the thermodynamic limit) is shown in Figure 9 at different energy per fold penalties as a function of the lattice row number. As can be inferred from that figure, that as the stiffness of chains increase, the interfacial layer, within which the dissipation of crystalline order takes place, becomes more and more thick already reaching the lattice half-height at an energetic bias of 2. As the energetic parameter increases even more, the crystalline order diminishes within the amorphous region. For a given chain stiffness (of strain energy) increasing $E_\eta$ is equivalent to a decrease in temperature causing a decrease in the overall chain mobility and thus "freezing" the microstructure. The annealing procedure described before allows for the structure to be frozen to the ones that correspond to a low energy one i.e. having a higher crystalline order (see Figure 12 for an example of possible structures).

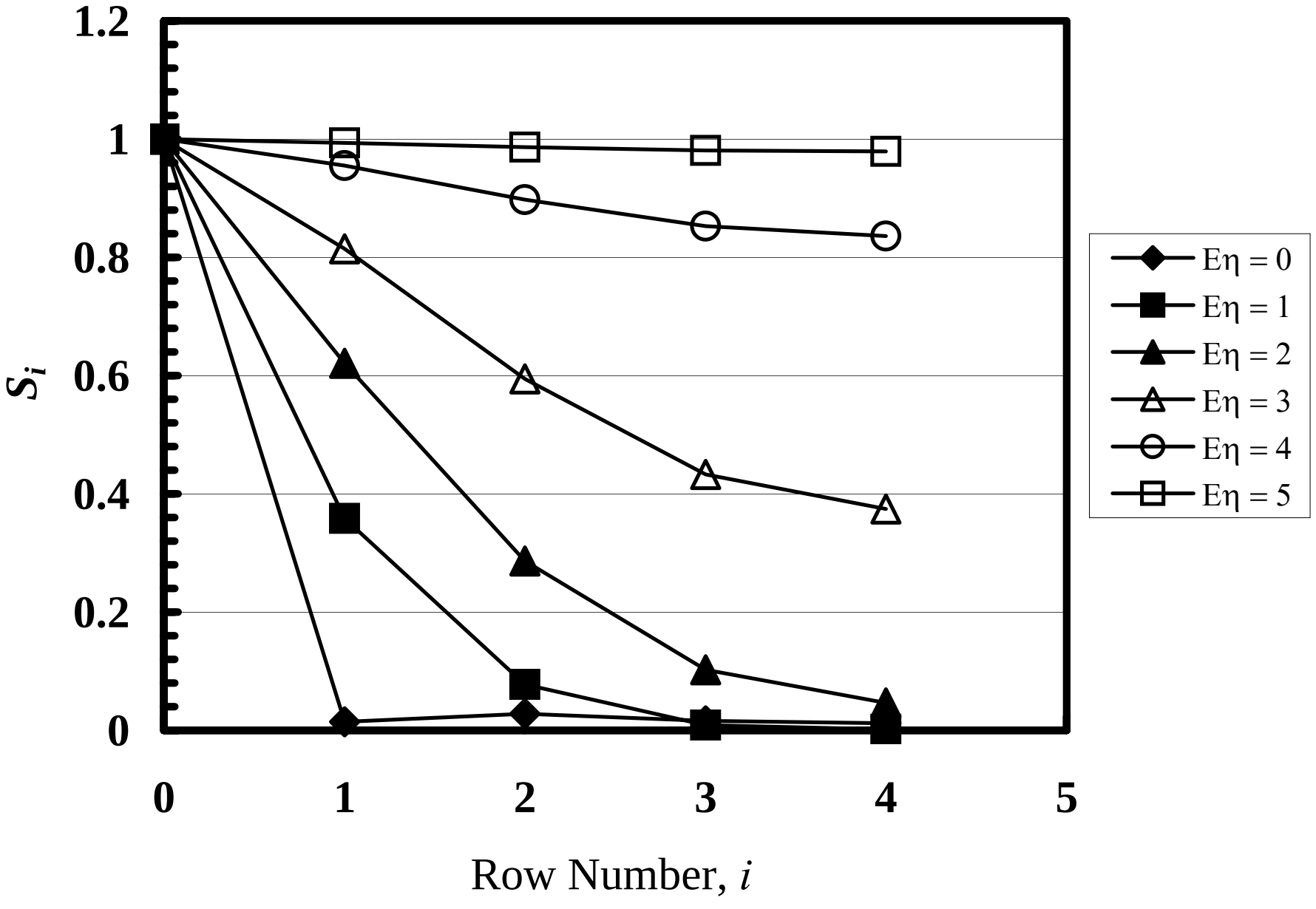

**Figure 9.** Order parameters (estimated at the thermodynamic limit) at different energy per fold penalties as a function of row number.

Another conformational statistics is shown in Figure 10, where the average length of bridges, as estimated at thermodynamic limit is shown as a function of the energy per fold penalty. Similarly in Figure 11, the average length of loops, estimated at thermodynamic limit is shown as a function of the energy per fold penalty, with a few sample microstates sampled at $E_\eta$ =5 shown in Figure 12, while in Figure 13 we evaluated the fraction of chains forming bridges and loops, as functions of energy per fold penalties.

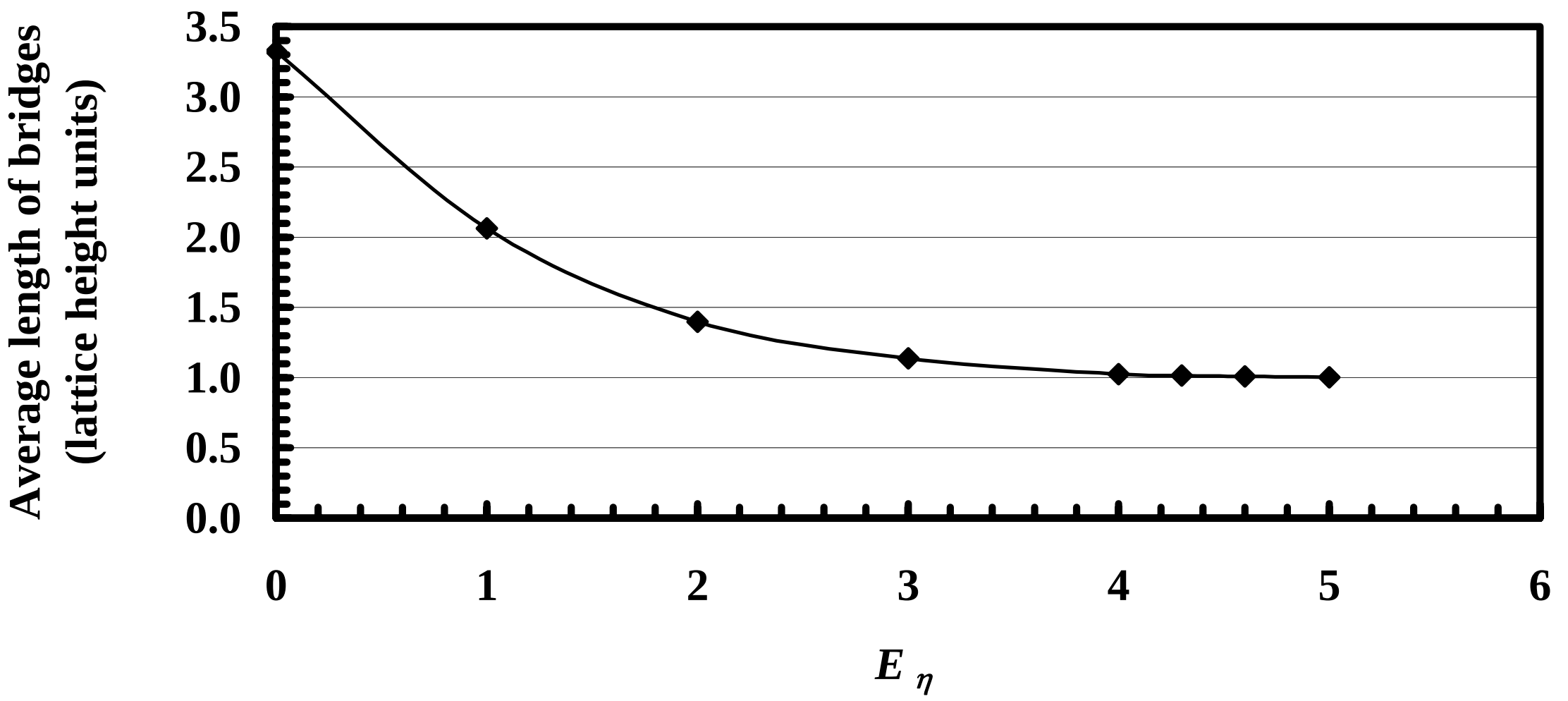

**Figure 10.** Average length of bridges estimated at thermodynamic limit as a function of energy per fold penalty.

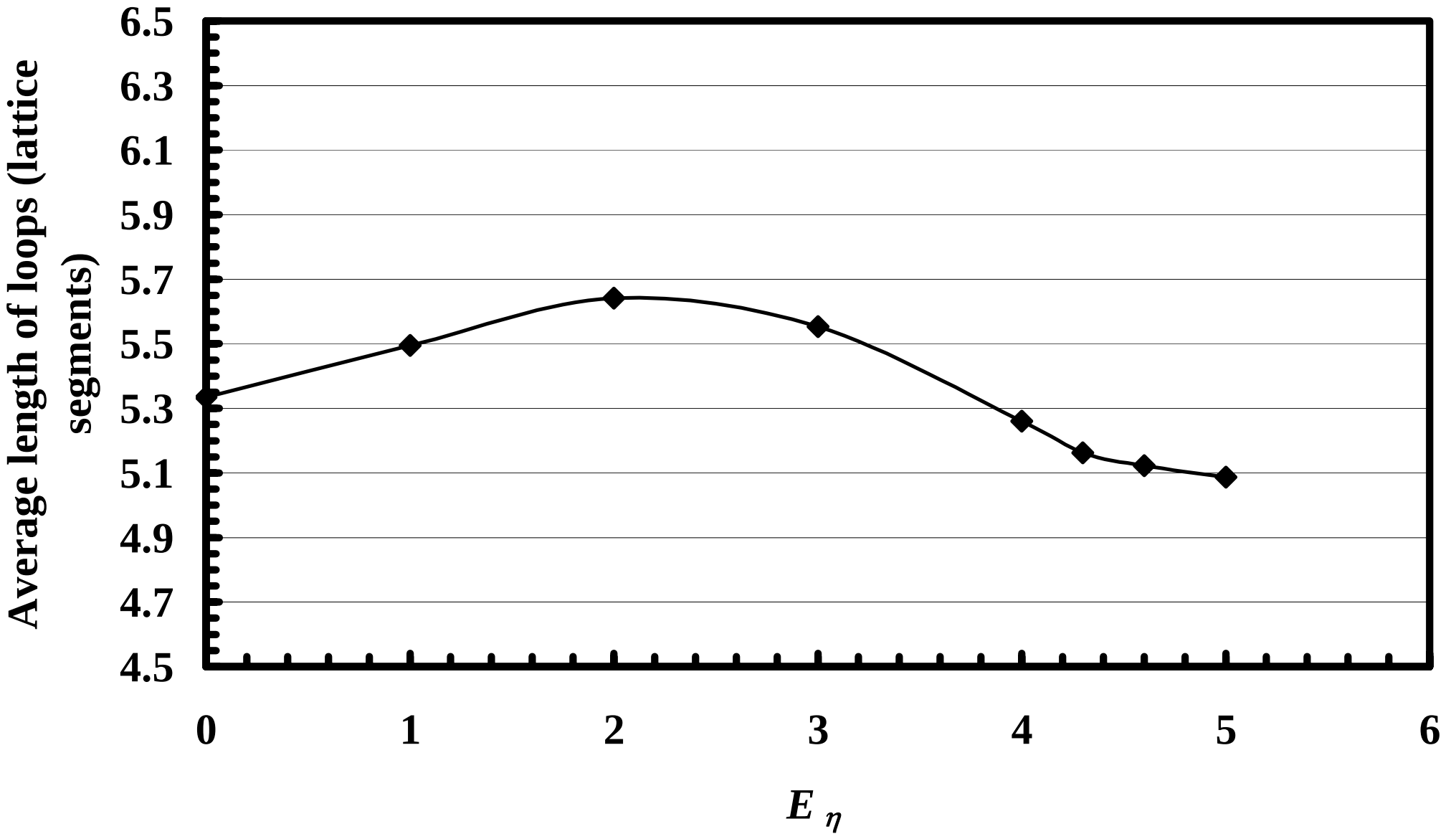

**Figure 11.** Average length of loops estimated at thermodynamic limit as a function of energy per fold penalty.

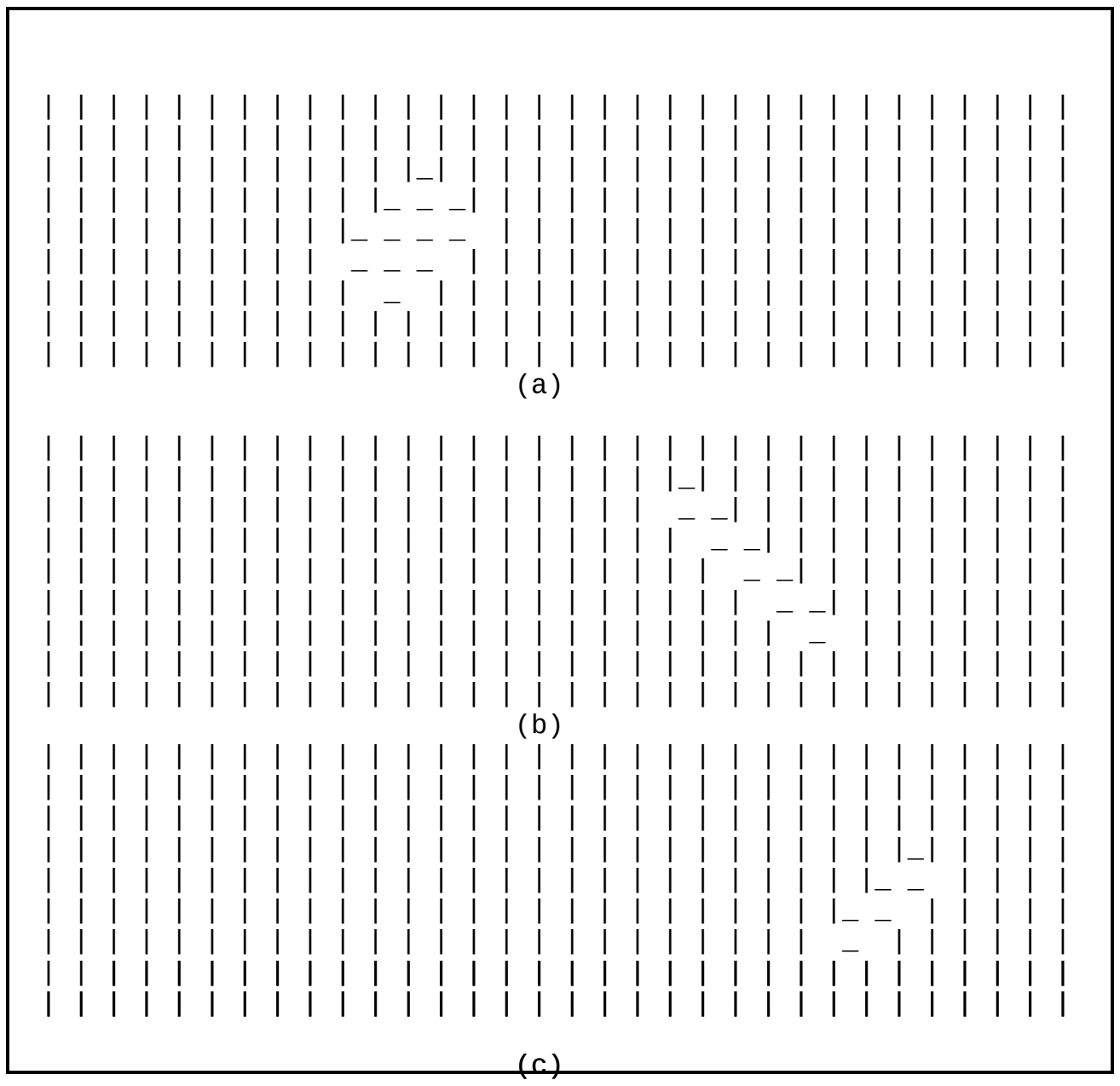

**Figure 12**. A few sample microstates of 8x32 lattice sampled at $E_\eta = 5$

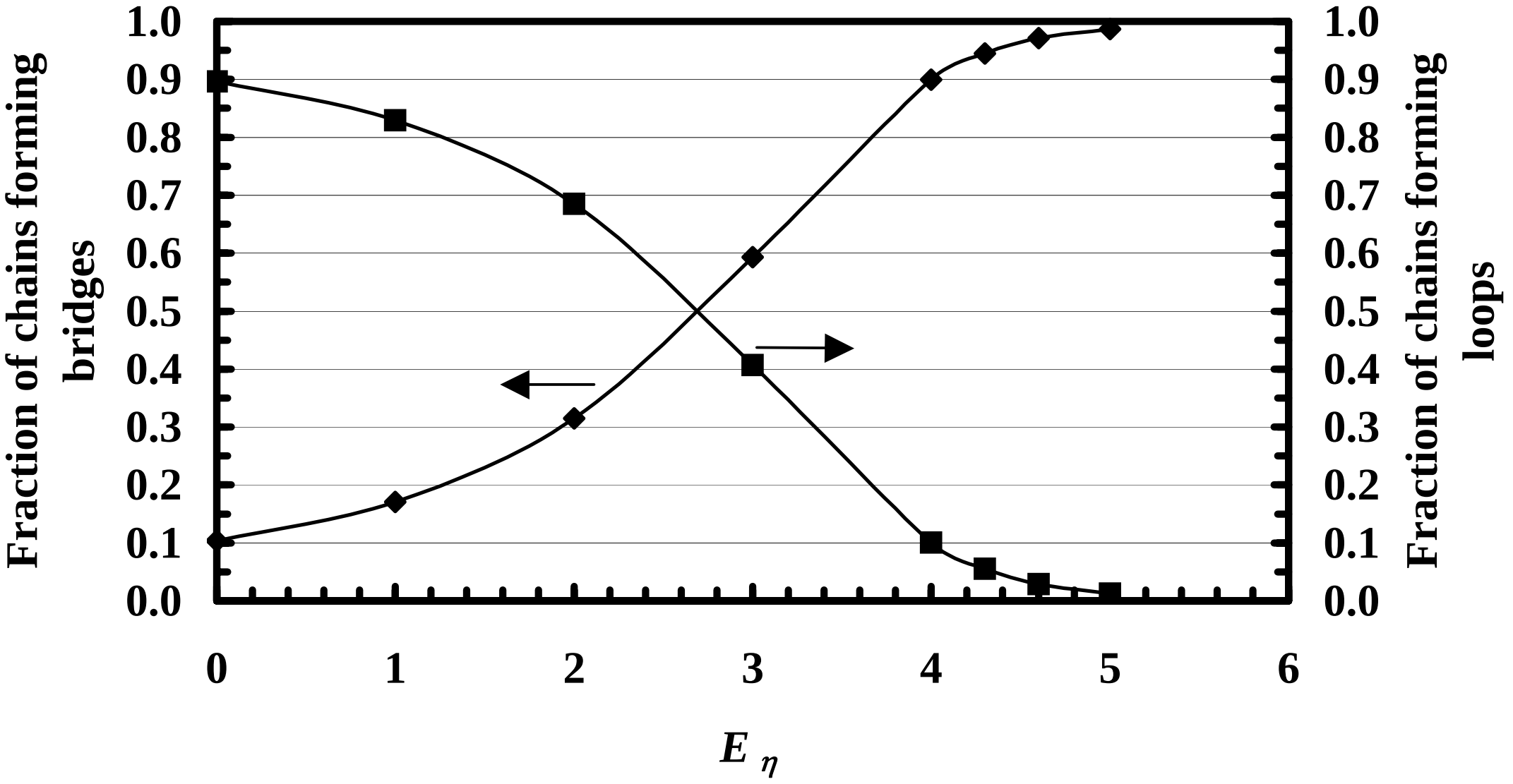

**Figure 13.** Fraction of chains forming bridges and loops estimated at thermodynamic limit as a function of energy per fold penalties.

The effect of energetic bias can easily be deduced from these figures. As the chain tight folding is penalized more and more, the length of bridges asymptotically tends to the interlamellar spacing as seen in Figure 10. Similarly, the average length of loops is seen to converge to a value close to 5.1 lattice segments. For an explanation of that limit, it is useful to look at Figure 12. There a few examples of microstates of an 8x32 lattice are sampled at $E_\eta$ =5. All the microstates shown have 2 tight folds and as such have high energetic penalties associated with them. However, there are a large number of such microstates (for example those generated from the ones shown in Figure 12 by shifting columns and/or by mirror imaging) and hence they are entropically favored. This presents a picture where a single imperfection like bending of a polymer chain in perfectly ordered state compels even stiff chains to form loops.

The fraction of chains in loops decreases monotonically, while the fraction chains forming bridges increases almost reaching a value of 1.0 at $E_\eta$ =5 implying that the lattice is almost crystallized (see Figure 13). Also note that these fractions add up to 1 indicating that a polymer chain is either a bridge or a loop thus negating the possibility of presence of an infinite chain that completely remains within the interlamellar amorphous regime.

### 5.2 Comparison with predictions based on gambler's ruin method

Gambler's ruin methods have been used widely in the past (see for example[8]) to predict chain conformational statistics. However, the theory is simple and is amenable to analytical analysis, it fails to accommodate excluded volume effects. Using the results from the classical gambler's ruin problem, it can be shown that the average lengths of bridges, $<B>$, and loops, $<L>$, in the 2-D case, are given by:

$$< B > \ = \frac{2(M^2 + 2M)}{3} + 1 \quad , \qquad \ldots 1$$

$$< L > \ = \frac{4M}{3} + 1 \quad , \qquad \ldots 2$$

where $M$ is the thickness of the amorphous region which equals $L\text{-}1= 7$ for a lattice with $8$x$N_0$ size. Using the above formula, the average length of bridges of an $8$x$N_0$ size lattice comes out to be 6.143 lattice height units and the average length of loops is predicted to be 10.333 lattice segments. These values are way too high compared to our simulation predictions at $E_\eta$ = 0. This shows the serious inadequacy of the random walk based models in handling excluded volume effects at least in 2-D square lattices.

### 5.3 Estimation of thermodynamic quantities for the amorphous region

Prediction of thermodynamic quantities is quintessential to perform any crystallization studies in semicrystalline polymers. Here we show the capability of our Monte Carlo scheme in conjunction with the stochastic enumeration scheme[1] to get quantitatively meaningful predictions of the absolute conformational entropy and the free energy of the amorphous interlamellar region.

In our model the free energy per node of the lattice, can be represented from principles of statistical thermodynamics in terms of the partition function Π as:

$$A = -k_B T \ln \Pi, \ \Pi = \sum_{i=1}^{\Omega} \exp(-E_\eta \times N_{tight\ folds_i}) \quad , \qquad \ldots 3$$

where Ω is the total number of microstates. The free energy has contributions from the internal energy, which accounts for the steric energy penalties due to intra-chain tight folding of polymer chains and from the conformational entropy. For completely flexible chains ($E_\eta$ = 0), the contribution due to internal energy is zero and the only contribution to free energy is due to the entropy of the polymer chains. Thus from equation 3 at $E_\eta$=0,

$$-\frac{A}{k_B T} = \frac{S}{k_B} = \ln \Omega \quad , \qquad \ldots 4$$

where Ω, the total number of microstates, is estimated using the stochastic enumeration scheme which is discussed in detail in ref. [1]. Thus both the free energy and the entropy can be estimated at $E_\eta$ = 0. Furthermore, from equation 3 the derivative of $A$ with respect to $E_\eta$ can be evaluated as:

$$\frac{\partial(-A/k_B T)}{\partial E_\eta} = -\overline{N_{tight\ folds}} \quad , \qquad \ldots 5$$

where $\overline{N_{tight\ folds}}$ is the average number of tight folds per microstate. Thus with the derivative known, an integration can be performed to evaluate the free energy as function of the energetic penalty factor.

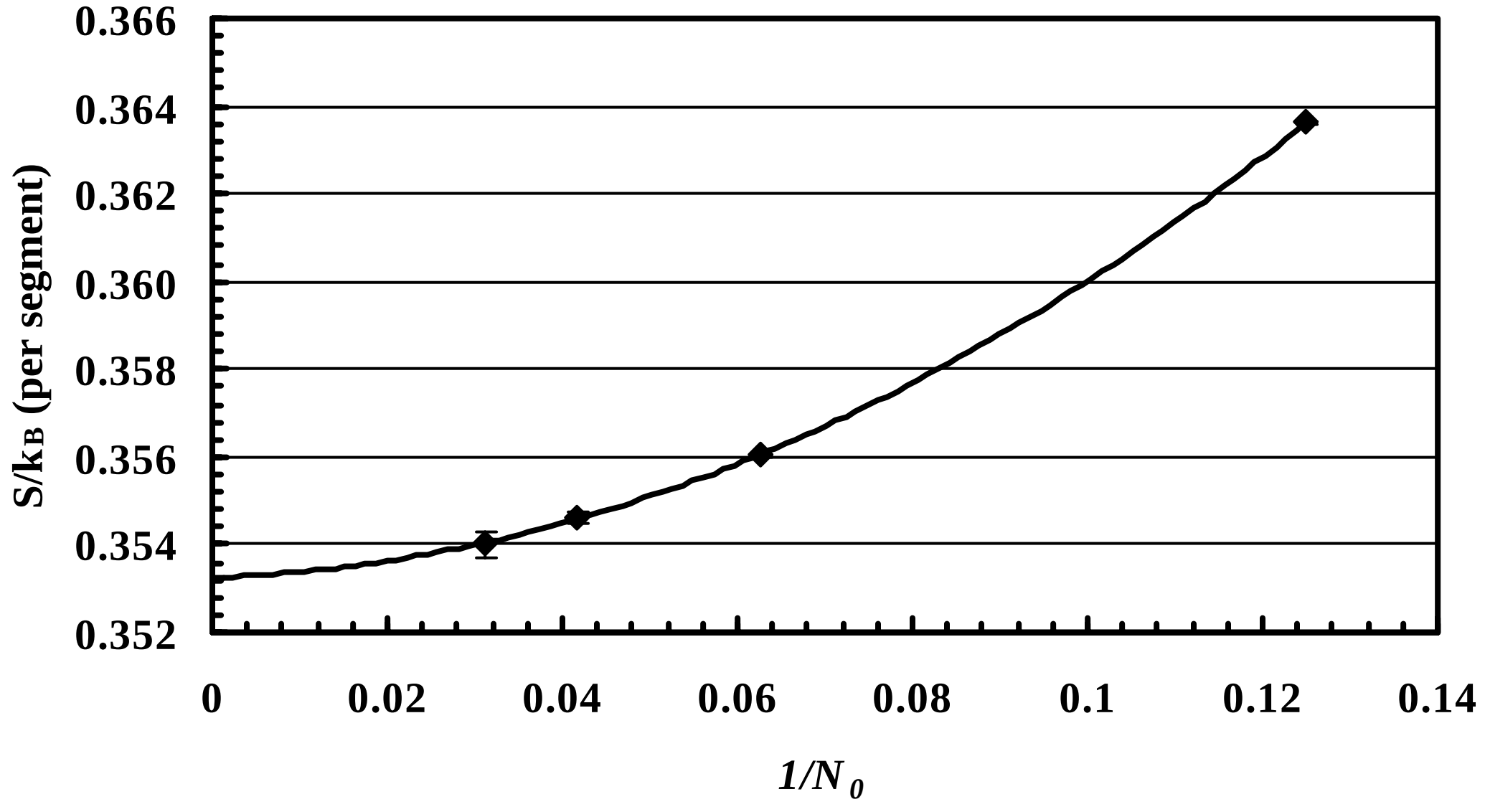

**Figure 14.** Conformational entropy per segment of polymer chain for different lattice sizes and extrapolated to the thermodynamic limit.

Using these derivations, the conformational entropy normalized to the number of segments in the lattice was determined for different lattice sizes and was extrapolated to the thermodynamic limit as shown in Figure 14. The value converges to 0.3533. This is compatible to the normalized entropy per segment for a single Hamiltonian walk on a 2-D Manhattan lattice (square lattice with alternating vertical and horizontal orientations), given as G/π ( approximately 0.29156…), where G is the Catalans constant: $G = 1 - \frac{1}{3^2} + \frac{1}{5^2} - \frac{1}{7^2} + .....$ (see ref. [9]). The literature result provides an absolute lower bound for the reported entropy given its restriction to Manhattan lattices. The close agreement is encouraging for the validity of our approach which, in fact, is the only approach providing such quantitatively significant information for general lattices which are not amenable to analytical analysis.

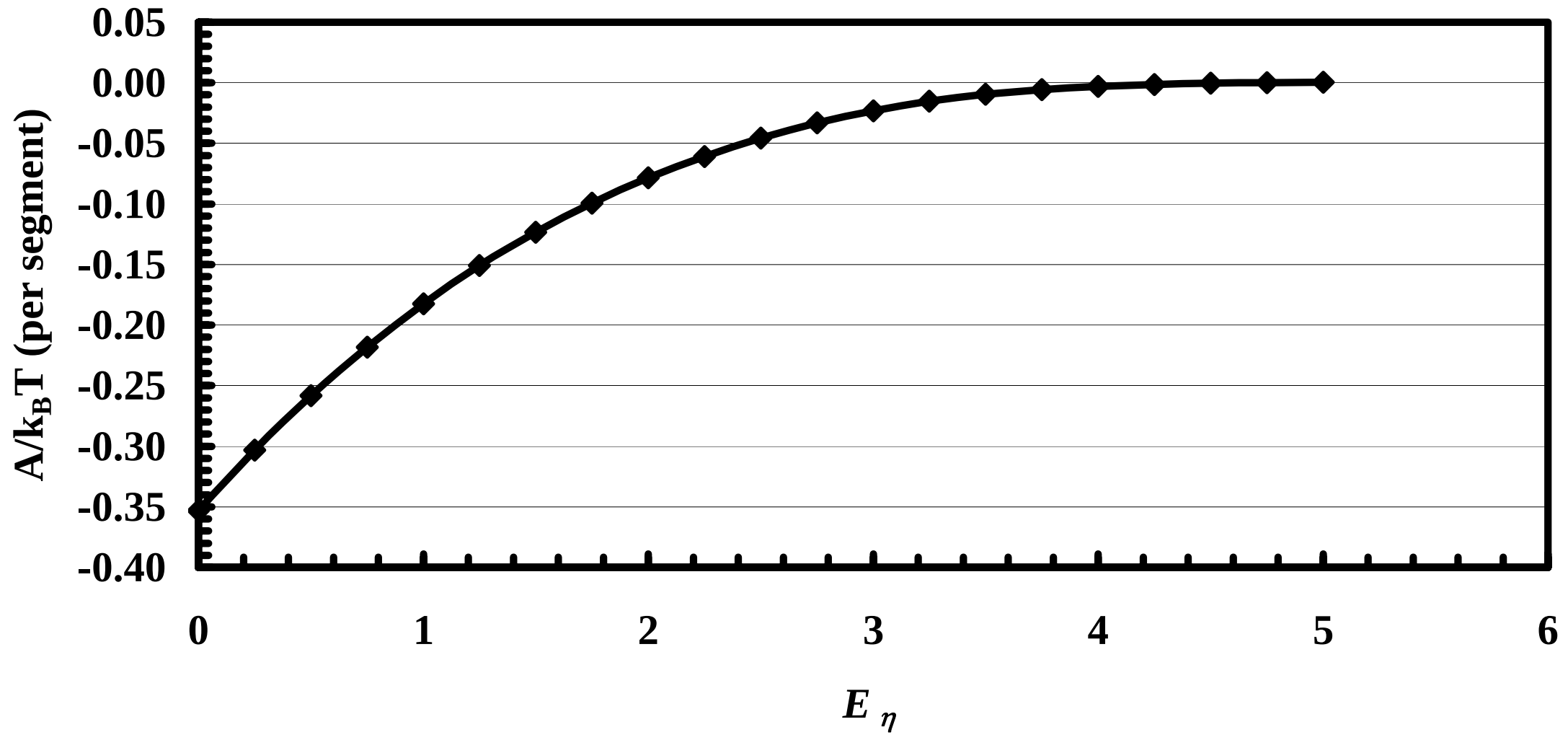

**Figure 15.** Free energy per segment of polymer chain in the interlamellar amorphous estimated at the thermodynamic limit and represented as a function of the parameter $E_\eta$.

Carrying out the integration with respect to $E_\eta$, the free energy was determined as a function of the energetic penalty parameter $E_\eta$ as shown in Figure 15. It is evident that the free energy asymptotically approaches to 0, as the stiffness of chains increases, indicating for 2-D lattices the topological constraints of the crystalline to amorphous interface are so strong so that to propagate the interface well within the bulk when significant steric hindrance (high $E_\eta$) is present.

## 6 Conclusions

In this work we showed results from the application of robust and efficient methodologies for the evaluation of quantitative information for the thermodynamics and conformational statistics of the amorphous interlamellar region of semicrystalline polymers using our novel lattice–subdivision based Monte Carlo scheme as modeled by a 2-D square lattice model. Although the topological bias induced by the two dimensions used in the present analysis is in all likelihood too high for the results of the present analysis to be directly applicable to semicrystalline lamellar polymer morphologies (that may have to be delegated for the 3-D analysis that follows in the third part of this series[10]) the value of this work is still significant. First we demonstrated the utility of a systematic methodology in order to develop a tool for the quantitative investigation of macromolecular structures and thermodynamics with strict error bounds. Second, the results are of direct importance in situations where you indeed have a 2-D morphology such as adsorption of macromolecules on surfaces and interfaces.

In particular, we showed the intricacy of the problem applied to 2-D lattices due to lattice size and space filling criterion. Without the tight error bounds imposed on the analysis, the very essential to the problem fact of the existence of two separate and mutually exclusive populations, one with an even number of horizontal segments in the bridges and the other one with odd, would have been missed. Furthermore, two advanced adaptations of the Monte Carlo scheme, simulated annealing and parallel tempering were developed to investigate lattices of larger sizes effectively at high energetic biases avoiding local energy minima trapping. Both the schemes were seen to be consistent with respect to each other and were used for the investigation of size and energy effects. Finally we show that our Monte Carlo scheme gives us consistent and quantitatively significant estimates of chain conformational and thermodynamic variables in the thermodynamic limit.

## Acknowledgments

The authors will like to acknowledge the National Science Foundation (award DMI 9978656 in materials processing and manufacturing) for funding this research.

## References

[1] J.A. Kulkarni, J. Mukherjee, R.C. Snyder, T. King, A. N. Beris, *A Novel Approach for Lattice Simulations of Polymer Chains in Dense Amorphous Polymer Systems: Method Development and Validation with 2-D Lattices,* submitted to ArXiv, **2008.**
[2] D.P. Landau, K. Binder, "*A guide to Monte Carlo Simulations for Statistical Physics*", Cambridge University Press, **2000**.
[3] M.E.J. Newman, G.T. Barkema, "*Monte Carlo Methods in Statistical Physics*", Oxford University Press, **1999**.
[4] E. Shakhnovich, G. Farztdinov, A.M.Gutin, M. Karplus, Phys. Rev. Lett., **1991**, 67:1665-1668.
[5] J.C. Schön, J.Phys. Chem. A, **2002**, 106:10886-10892.
[6] R.H. Swendsen, J.S. Wang, Phys. Rev. Lett., **1987**, 58:86-88.
[7] B.A. Berg, T. Neuhaus, Phys. Rev. Lett., **1992**, 68:9-12.
[8] C.M. Guttman, E.A. DiMarzio, J.D. Hoffman, Polymer, **1981**, 22:1466-1477.
[9] B. Duplantier and H. Saleur, Nuclear Phys., **1987**, B290:291–326.
[10] S. Wilson, J. Mukherjee, A.N.Beris**,** *A Novel 3-D Lattice Monte Carlo Method for Dense Semicrystalline Polymers*, to be submitted to Macromolecules, **2008**.